\documentclass[twocolumn,english,showpacs,preprintnumbers]{revtex4}
\usepackage[T1]{fontenc}
\usepackage[latin9]{inputenc}
\usepackage{amsmath}
\usepackage{graphicx}
\usepackage{amssymb}
\usepackage{esint}

\makeatletter
\@ifundefined{textcolor}{}
{%
 \definecolor{BLACK}{gray}{0}
 \definecolor{WHITE}{gray}{1}
 \definecolor{RED}{rgb}{1,0,0}
 \definecolor{GREEN}{rgb}{0,1,0}
 \definecolor{BLUE}{rgb}{0,0,1}
 \definecolor{CYAN}{cmyk}{1,0,0,0}
 \definecolor{MAGENTA}{cmyk}{0,1,0,0}
 \definecolor{YELLOW}{cmyk}{0,0,1,0}
 }

\usepackage{amsfonts}

\usepackage{babel}

\begin{document}

\title{Single Impurity In Ultracold Fermi Superfluids}

\author{Lei Jiang$^{1}$, Leslie O. Baksmaty$^{1}$, Hui Hu$^{2}$, Yan Chen$^{3}$
and Han Pu$^{1}$}

\affiliation{$^{1}$Department of Physics and Astronomy, and Rice Quantum Institute,
Rice University, Houston, TX 77251, USA \\
 $^{2}$ARC Centre of Excellence for Quantum-Atom Optics, Centre
for Atom Optics and Ultrafast Spectroscopy, Swinburne University of
Technology, Melbourne 3122, Australia \\
 $^{3}$3Department of Physics, State Key Laboratory of Surface Physics and Laboratory of Advanced Materials, Fudan University,
Shanghai, 200433, China }
\begin{abstract}
The role of impurities as experimental probes in the detection of
quantum material properties is well appreciated. Here
we study the effect of a single classical magnetic impurity in trapped
ultracold Fermi superfluids. Depending on its shape and
strength, a magnetic impurity can induce single or multiple
mid-gap bound states in a superfluid Fermi gas. The multiple mid-gap
states could coincide with the development of a Fulde-Ferrell-Larkin-Ovchinnikov (FFLO) phase within
the superfluid. As an analog of the Scanning Tunneling Microsope,
we propose a modified RF spectroscopic method to measure the local
density of states which can be employed to detect these states and other quantum phases of cold atoms. A key result
of our self consistent Bogoliubov-de Gennes calculations is that
a magnetic impurity can controllably induce an FFLO state at currently
accessible experimental parameters.
\end{abstract}

\date{\today }


\maketitle Trapped ultra-cold gases represent a many-body quantum
system amenable to selective experimental control, and possess some
notable advantages in comparision with conventional quantum
many-body systems such as solid state materials. With a view to some
of these properties such as the accurate control of inter-particle
interaction or density and the use of laser light to simulate
external potentials, we anticipate important contributions from
these new experimental systems through the study of impurities. Due
to their unavoidable and ubiquitous presence in real materials, the
effects of impurities constitute an important and sometimes
frustrating issue in condensed matter physics. However, under many
circumstances, impurities, rather than representing a nuisance,
serve useful purposes such as the detection of quantum effects
\cite{zhu06,Alloul09}. Single impurities have been employed in the
detection of superconducting pairing symmetry within unconventional
superconductors \cite{Mackenzie98} and to demonstrate Friedel
oscillations \cite{Sprunger97}. In strongly correlated systems, they
may be used to pin one of the competing orders \cite{Millis03}. Even
though cold atom systems are intrinsically clean, the effects of
impurities may be simulated by employing laser speckles or
quasiperiodic lattices \cite{Modugno10}. Controllable manipulation
of individual impurities in cold atom systems can also be realized
using off-resonant laser light or another species of atoms/ions
\cite{Zipkes10,demler,Stringari}. Such impurities can be either
localized or extended and either static or dynamic. The
unprecedented access to accurately tune these artificial impurities,
provide an exciting possibility to probe and manipulate the
properties of cold atoms.

In this Letter, we demonstrate this possibility using a single classical
static impurity in an \textit{s}-wave Fermi superfluid. By `classical'
we refer to the treatment of the impurity as a scattering potential
which has no internal degrees of freedom. We focus on a magnetic
impurity which scatters each spin species differently. From our self-consistent
Bogoliubov-de Gennes calculations we show for the first time in a trapped three-dimensional (3D) geometry
that the long sought Fulde-Ferrell-Larkin-Ovchinnikov (FFLO) phase, which supports many mid-gap bound states, may be induced through
such an impurity at experimentally accessible parameters.
Futhermore, we propose that these bound states can be probed using a
modified radio-frequency (RF) spectrosocpy technique that is the analog
of the widely used scanning tunneling microscope (STM) in solid state and that this can
serve as a powerful general tool in probing and manipulating quantum gases.

For computational simplicity, we focus on a one-dimensional (1D) system
and verify the essential physics at higher dimensions in later paragraphs.
Consider the following Hamiltonian at zero temperature,\begin{eqnarray}
H & = & \underset{\sigma=\uparrow,\downarrow}{\sum}\int dx\,\psi_{\sigma}^{\dag}\left[-\frac{\hbar^{2}}{2m}\frac{d^{2}}{dx^{2}}-\mu_{\sigma}+V_{T}\right]\psi_{\sigma}\notag\\
 &  & +g\int dx\,\psi_{\uparrow}^{\dag}\psi_{\downarrow}^{\dag}\psi_{\downarrow}\psi_{\uparrow}+\underset{\sigma=\uparrow,\downarrow}{\sum}\int dx\,\psi_{\sigma}^{\dag}U_{\sigma}\psi_{\sigma},\label{eq: Hamiltonian}\end{eqnarray}
 where $\psi_{\sigma}^{\dag}(x)$ and $\psi_{\sigma}(x)$ are, respectively,
the fermionic creation and annihilation operators for spin species
$\sigma$. $V_{T}(x)$ is a harmonic trapping potential and $g$ is
the strength of the inter-atomic interaction. In this work, we take
$g$ to be small and negative so that the system is a superfluid at
low temperatures. The last term of the Hamiltonian describes the effect
of the impurity which is represented by a scattering potential, $U_{\sigma}(x)$.
For non-magnetic impurity$U_{\uparrow}(x)=U_{\downarrow}(x)$; while
for magnetic impurity, $U_{\uparrow}(x)=-U_{\downarrow}(x)$. Note
that a general impurity potential can be decomposed into a sum of
magnetic and non-magnetic parts. Here we focus on magnetic impurities
which can be either localized or extended.

\begin{figure}[ptb]

\begin{center}
\includegraphics[width=3.3in]{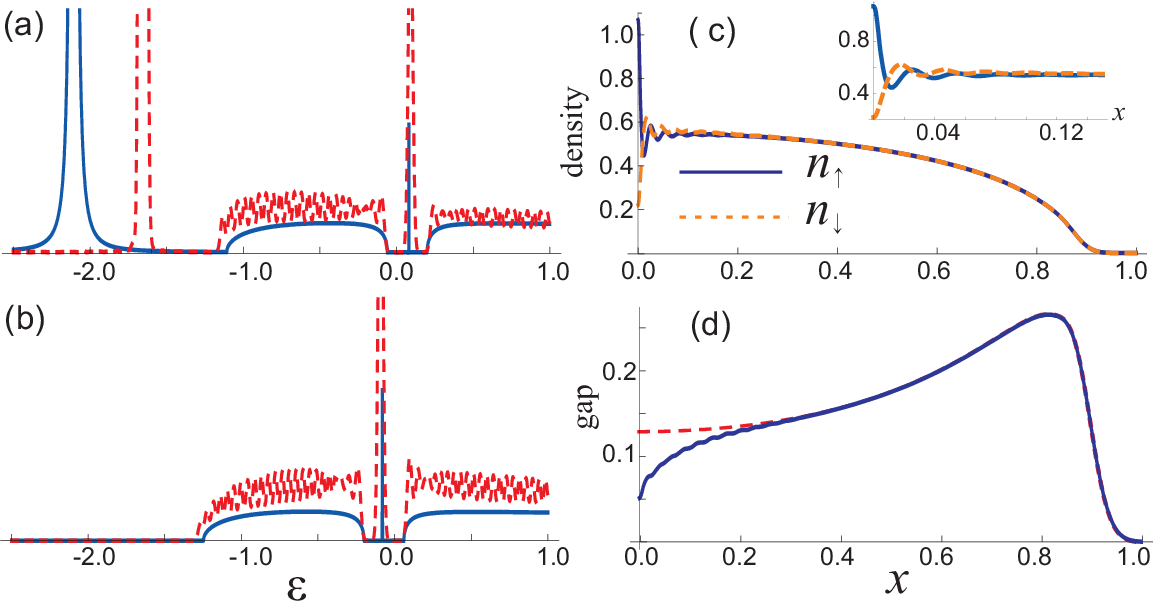}\caption{(a) Density of states for spin up atoms. (b) Density of states for
spin down atoms. (c) Density profiles for both spin species. (d) Gap
profile. In (a) and (b) solid and dashed lines represent results obtained
using the $T$-matrix and BdG method, respectively. The dashed curve
in (d) is the gap profile without the impurity. For all plots, $N_\uparrow=N_\downarrow=50$,
and $u=-0.02E_{F}x_{TF}$, where $E_{F}$ is the Fermi energy and
$x_{TF}=\sqrt{N}a_{{\rm ho}}$ is the Thomas-Fermi radius of the non-interacting
system. The harmonic oscillator length and Thomas Fermi density at
the origin are defined by $a_{{\rm ho}}=\sqrt{\hbar/(m\omega_{0})}$
and $n_{0}=2\sqrt{N}/(\pi a_{{\rm ho}})$. The dimensionless interaction
parameter $\gamma=-mg/(\hbar^{2}n_{0})=1.25$. The units for density,
energy and length are $n_{0}$, $E_{F}$ and $x_{TF}$, respectively. }
\label{Fig:magimp}
\end{center}

\end{figure}

{\em Localized impurity ---} Let us first consider a localized
impurity with $U_{\sigma}(x)=u_{\sigma}\delta(x)$. If we restrict
ourselves to the vicinity of the impurity, we may neglect
the trapping potential and use the $T$-matrix formalism \cite{Hirschfeld86}.
As a result of the $\delta$-function impurity potential, the $T$-matrix
is momentum independent and analytical results can be obtained. The
full Green's function $G$ is related to the bare (i.e., in the absence
of the impurity) Green's function $G{}_{0}$ and the $T$-matrix in
the following way: \begin{equation}
G(k,k^{\prime},\omega)=G_{0}(k,\omega)\delta_{kk^{\prime}}+G_{0}(k,\omega)T(\omega)G_{0}(k^{\prime},\omega),\end{equation}
 where $\omega$ is the frequency, $k$ and $k'$ represent the incoming
and outgoing momenta in the scattering event, respectively. For the
$s$-wave superfluid, we have: \begin{equation}
G_{0}(k,w)=\frac{\omega\sigma_{0}+(\epsilon_{k}-\tilde{\mu})\sigma_{3}-\Delta\sigma_{1}}{\omega^{2}-(\epsilon_{k}-\tilde{\mu})^{2}-\Delta^{2}},\end{equation}
 where $\epsilon_{k}=\hbar^{2}k^{2}/(2m)$, $\sigma_{i}$'s are the
Pauli matrices ($\sigma_{0}$ is the identity matrix) and $\Delta$
is the $s$-wave pairing gap. Here the effective chemical potential,
$\tilde{\mu}=\mu-gn(x)$, includes the contribution from the Hartree
term, where $n(x)$ is the local density for one spin species. For
magnetic impurity, we take $u=u_{\uparrow}=-u_{\downarrow}$and the
$T$-matrix is given by: \begin{equation}
T^{-1}(\omega)=u^{-1}\sigma_{0}-{\sum_{k}}G_{0}(k,\omega)\,,\label{tnon}\end{equation}
 while for non-magnetic impurity with $u=u_{\uparrow}=u_{\downarrow}$,
and the corresponding $T$-matrix has the same form as in Eq.~(\ref{tnon})
with $\sigma_{0}$ replaced by $\sigma_{3}$. From the full Green's
function, one can immediately obtain the local density of states (LDOS)
at the impurity site as: \begin{equation}
\rho(\epsilon)=-\frac{1}{\pi}\sum_{k,k'}{\rm Im}\left[G(k,k',\epsilon+i0^{+})\right]\,.\end{equation}
The solid lines in Fig.~\ref{Fig:magimp}(a),(b) display the LDOS
at the magnetic impurity site for the two spin species obtained using
the $T$-matrix method. Here the impurity potential is attractive
(repulsive) for spin-up (down) atoms which creates a resonant state
below the Fermi sea for spin up atoms manifested by the peak near
$\epsilon=-2E_{F}$ in Fig.~\ref{Fig:magimp}(a). 
As the strength of the impurity potential $|u|$ increases, the resonant
state will move deeper below the Fermi sea. Besides this resonant state,
both $\rho_{\uparrow}(\epsilon)$ and $\rho_{\downarrow}(\epsilon)$
exhibit an additional peak near $\epsilon=0$, which signals the presence
of a mid-gap bound state \cite{yu65}. In the limit of weak interaction,
the position of the mid-gap bound state is given by the $T$-matrix
method as: \begin{equation}
E_{0}=\pm\Delta\,\frac{1-(u\pi\rho_{0}/2)^{2}}{1+(u\pi\rho_{0}/2)^{2}}\,,\end{equation}
 where $\rho_{0}$ is the density of states at the Fermi sea, and
the $+$ ($-$) sign refers to the spin-up (-down) component. The
mid-gap bound state is thus located outside the band and inside the
pairing gap. As the strength of the impurity $|u|$ increases, the
mid-gap moves from the upper gap edge to the lower gap edge for spin-up
component and moves oppositely for spin-down component.

To confirm that these results still hold when a trapping potential
is present, as is always the case in the experiment, we add a harmonic
potential $V_{T}=m\omega_{0}^{2}x^{2}/2$ to the system
and diagonalize the Hamiltonian using the Bogoliubov-de Gennes (BdG)
method \cite{de Gennes66,Hu07,BdG}: \begin{eqnarray}
 &  & \left[\begin{array}{cc}
H_{\uparrow}^{s}-\mu_{\uparrow} & -\Delta\\
-\Delta^{\ast} & -H_{\downarrow}^{s}+\mu_{\downarrow}\end{array}\right]\left[\begin{array}{c}
u_{\eta}\\
v_{\eta}\end{array}\right]=E_{\eta}\left[\begin{array}{c}
u_{\eta}\\
v_{\eta}\end{array}\right]\,,\end{eqnarray}
 where $H_{\sigma}^{s}=-(\hbar^{2}/2m)d^{2}/dx^{2}+V_{T}+U_{\sigma}+gn_{\bar{\sigma}}$,
$n_{\uparrow}=\sum_{\eta}|u_{\eta}|^{2}\Theta(-E_{\eta})$, $n_{\downarrow}=\sum_{\eta}|v_{\eta}|^{2}\Theta(E_{\eta})$
and $\Delta=-g\sum_{\eta}u_{\eta}v_{\eta}^{*}\Theta(-E_{\eta})$,
with $\Theta(\cdot)$ being the unit step function. The BdG equations
above are solved self-consistently using a hybrid method whose details
can be found in Ref.~\cite{Hu07}. Once the solutions are found,
we can calculate the LDOS at any points in space as $\rho_{\uparrow}(\epsilon)=\sum_{\eta}|u_{\eta}|^{2}\delta(\epsilon-E_{\eta})$
and $\rho_{\downarrow}(\epsilon)=\sum_{\eta}|v_{\eta}|^{2}\delta(\epsilon+E_{\eta})$.
In practice, the $\delta$-function in the expression of the LDOS
is replaced by a Gaussian with a small width of 0.02 $E_{F}$.

The dashed line in Fig.~\ref{Fig:magimp}(a),(b) represents the LDOS
at the magnetic impurity site ($x=0$) calculated using the BdG method.
The agreement with the $T$-matrix method is satisfactory. The remaining
discrepancies such as in the position of the resonant state below
Fermi sea can be understood by considering that the $T$-matrix method
neglects the trapping potential and is not fully self-consistent:
the values of the chemical potentials, densities and pairing gap used
in the $T$-matrix calculation are taken to be those from the BdG
result in the absence of the impurity. The density and gap profiles
of the trapped system are illustrated in Fig.~\ref{Fig:magimp}(c),(d).
Friedel oscillations with a spatial frequency close to $2k_{F}$ can
be seen in the density profiles near the impurity. The magnetic impurity
tends to break Cooper pairs, leading to a reduced gap size near the
impurity as can be seen from Fig.~\ref{Fig:magimp}(d).

{\em Detection of the mid-gap state ---} As we have seen above, the
mid-gap bound state induced by a magnetic impurity manifests itself
in the LDOS. In general the LDOS provides valuable information on
the quantum system and it is highly desirable to measure it
directly. Great dividends have been reaped in the study of high
$T_{c}$ superconductors where the scanning tunneling microscope
(STM), which measures the differential current proportional to the
LDOS, provides this function\cite{STM}. In ultra-cold Fermi gases,
radio-frequency (RF) spectroscopy \cite{Regal03,Gupta03,Grimm04}
could serve as an analogous tool. The RF field induces
single-particle excitations by coupling one of the spin species (say
$|\uparrow\rangle$ atoms) out of the pairing state to a third state
$|3\rangle$ which is initially unoccupied. In the experiment, the RF
signal is defined as the average rate change of the population in
state $|\uparrow\rangle$ (or state $|3\rangle$) during the RF pulse.
The first generation RF had low resolution and provided averaged
currents over the whole atomic cloud, which complicated
interpretation of the signal due to the inhomogeneity of the sample
\cite{stoof}. More recently spatially resolved RF spectroscopy which
provides {\em local} information has been demonstrated
\cite{Ketterle07}. Here we show that a modified implementation of
the spatially resolved RF spectroscopy can yield direct information
of the LDOS and hence can serve as a powerful tool in the study of
quantum gases.

To study the effect of the RF field, we make two additions to the
total Hamiltonian (\ref{eq: Hamiltonian}): \begin{eqnarray*}
H_{3} & = & \int dx\,\psi_{3}^{\dag}(x)\left[-\frac{\hbar^{2}}{2m}\frac{d^{2}}{dx^{2}}+V_{3}(x)-\nu-\mu_{3}\right]\psi_{3}(x),\\
H_{T} & = & \int dx\,[T\psi_{3}^{\dag}(x)\psi_{\uparrow}(x)+T\psi_{\uparrow}^{\dag}(x)\psi_{3}(x)],\end{eqnarray*}
 where $H_{3}$ represents the single-particle Hamiltonian of the
state 3 (we assume that atoms in state 3 do not interact with other
atoms), with $V_{3}$ being the trapping potential of the state and
$\nu$ the detuning of the RF field from the atomic transition, $H_{T}$
represents the coupling between state 3 and spin-up atoms. Since RF
photon wavelength is much larger than typical size of the atomic cloud,
the coupling strength $T$ can be regarded as a spatially invariant
constant. For weak RF coupling, one may use the linear response theory
\cite{torma00,torma04,levin05} to obtain the RF signal which is proportional
to $I(x)=\frac{d}{dt}\langle\psi_{3}^{\dag}(x)\psi_{3}(x)\rangle$.
Under the linear response theory, we have \[
I(x)\propto \int dx^{\prime}d\omega A_{\uparrow}(x,x^{\prime};\omega)A_{3}(x^{\prime},x,\omega+\mu_{\uparrow}-\mu_{3})f(\omega)\,,\]
 where $f(\omega)$ is the Fermi distribution function which reduces
to the step function at zero temperature, $A_{\alpha}$ is the spectral
function for state $\alpha$. As state 3 is non-interacting, we have
$A_{3}={\sum_{n}}\phi_{n}(x)\phi_{n}^{\ast}(x^{\prime})\delta(\omega+\mu_{\uparrow}+\nu-\epsilon_{n})$,
where $\phi_{n}$ and $\epsilon_{n}$ are the single-particle eigenfunctions
and eigen-energies of state 3, respectively. The key step in our proposal is that
in the case where $V_{3}$
represents an optical lattice potential in the tight-binding limit,
the dispersion of state 3 is proportional to the hopping constant
$t$ which decreases exponentially as the lattice strength is increased.
For sufficiently large lattice strength, we may therefore neglect
the dispersion of state 3 since the lowest band is nearly flat. In other
words, under such conditions, $\epsilon_{n}=\epsilon$ becomes an
$n$-independent constant. Consequently $A_{3}(x,x')\sim\delta(x-x')$.
In this limit and at zero temperature, the RF signal is then is directly related to the
LDOS as: \begin{equation}
I(x)\propto \rho_{\uparrow} (x,-\mu_{\uparrow}-\nu+\epsilon)\Theta(\mu_{\uparrow}+\nu-\epsilon)\,.\end{equation}
 and the spatially resolved RF spectroscopy becomes a direct analog
of the STM. A crucial point here is that {\em only} state 3 experiences the lattice
potential. We note here that a spin-dependent optical lattice selectively affecting
only one spin state has recently been realized in the lab of de Marco
\cite{marco}. The same technique can also be used to create magnetic
impurity potentials by external light field.

\begin{figure}
[ptb]
\begin{center}
\includegraphics[
width=3.2in
]%
{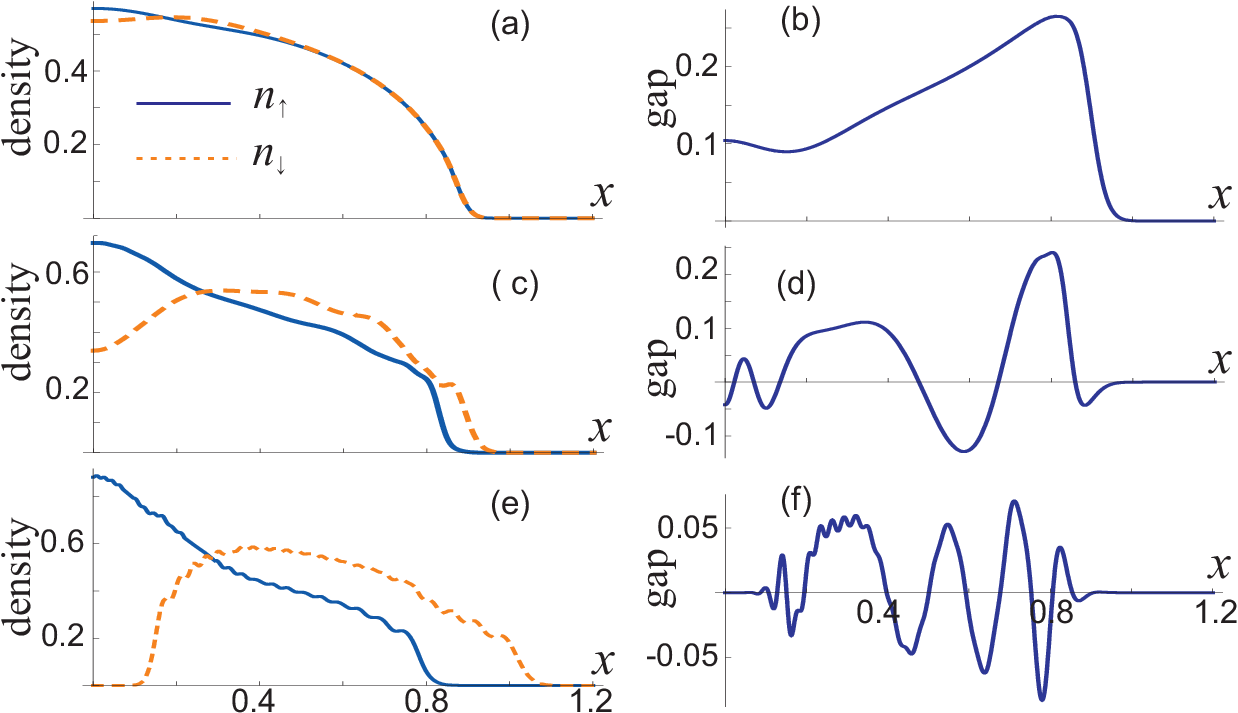}%
\caption{(Color online) Density (left panel) and gap (right panel) profiles of a trapped system under an extended Gaussian magnetic impurity potential. The width of the impurity potential is $a =0.2x_{TF}$, while the strength is $u=-0.12E_{F}x_{TF}$ for (a) and (b); $u=-0.4E_{F}x_{TF}$ for (c) and (d), and $u=-1.0E_{F}x_{TF}$ for (e) and (f).
Other parameters and units are the same as in Fig.~\ref{Fig:magimp}.}%
\label{Fig:FFLO}%
\end{center}
\end{figure}

{\em Extended impurity ---} Now we turn to Gaussian impurity potentials
with finite width $U_{\sigma}(x)=u_{\sigma}e^{-x^{2}/a^{2}}/(a\sqrt{\pi})$.
Since we obtain all of the previous (delta function) physics for narrow
widths, we focus on relatively wide potentials. Examples of the density
and gap profiles obtained from our BdG calculations are shown in Fig.~\ref{Fig:FFLO}.
For an extended impurity potential of sufficient width, the Friedel
oscillations are suppressed. Under appropriate conditions, the gap
profiles exhibit FFLO-like oscillations
\cite{FFLO}, which has recently received considerable attention in
studies of ultra-cold atoms \cite{randy,Orso,Demler}. In previous
experiments, polarized Fermi gas have been realized by preparing the
gas with an overall population imbalance. Here the magnetic impurity
breaks the local population balance and by tailoring the strength
and/or the width of the magnetic impurity, one is able to control
the magnitude of the population imbalance as shown in Fig.~\ref{Fig:FFLO}
which in turn controls the nature of the induced FFLO state. The impurity
therefore provides us with a controlled way to create FFLO state.

\begin{figure}
[ptb]
\begin{center}
\includegraphics[
width=3.3  in
]%
{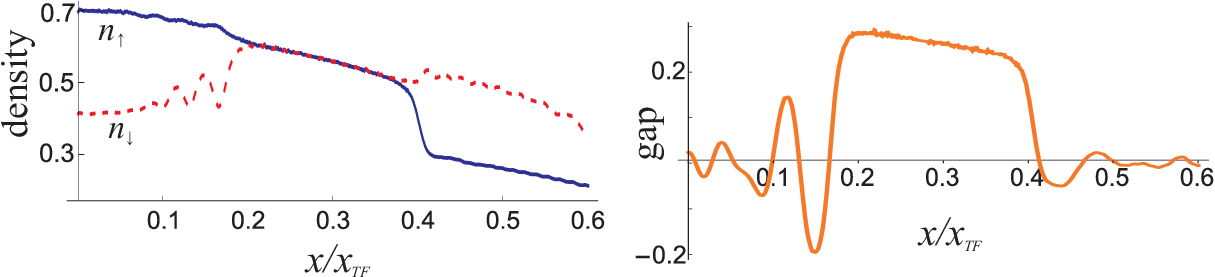}%
\caption{(Color online) Density (upper panel, in units of $(2E_F)^{3/2}/(6\pi^2)$) and gap (lower panel, in units of $E_F$) profiles along the $x$-axis of a 3D trapped system under an extended magnetic impurity potential. The impurity potential is uniform along the radial direction and has a Gaussian form with width $a=0.3x_{TF}$ along the $x$-axis. The strength of the impurity is $u=-0.07E_{F}x_{TF}$. The atom-atom interaction is characterized by the 3D scattering length $a_s$. Here we have used $1/(k_Fa_s) = -0.69$.
}%
\label{Fig:FFLO3D}%
\end{center}
\end{figure}

For simplicity, we have thus far focused on 1D systems. However, we
have verified that the essential physics is also valid in higher dimensions.
As an example, we illustrate in Fig.~\ref{Fig:FFLO3D} the effect
of an extended magnetic impurity in a 3D trapped system obtained by
solving the BdG equations \cite{BdG}. Here a total of 1100 atoms
are trapped in an elongated cylindrical trapping potential $V(r,x)=\frac{m}{2}(\omega_{\bot}^{2}r^{2}+\omega_{x}^{2}x^{2})$
with trap aspect ratio $\omega_{\bot}/\omega_{x}=50$. The magnetic
impurity centered at the origin, is radially uniform and has a Gaussian
profile along the axial direction ($x$-axis). From the density and
gap profiles shown in Fig.~\ref{Fig:FFLO3D}, one can easily identify
the induced FFLO regions both near the center and the edge of the
trap. In particular, the density oscillations in the spin-down component
near trap center may be used as a signature of the FFLO state.

In conclusion, we have investigated the effects of a single classical
magnetic impurity on a neutral fermionic superfluid. We show that a magnetic impurity can be used to manipulate novel quantum states in a Fermi gas. For example, it will induce a mid-gap bound state inside the
pairing gap for both spin species. We have proposed an STM-like scheme
based on the modified spatially resolved RF spectroscopy to measure the local density
of states, from which the mid-gap bound states can be unambiguously
detected. As different quantum phases of cold atoms will manifest thems`elves in their distinct LDOS, we expect this method will find important applications beyond what is proposed here and become an invaluable tool in the study of quantum gases. Finally, by considering an extended impurity potential in both 1D and 3D systems,
we demonstrate the realization of the still unobserved FFLO phase in a controlled
manner.

Interesting future directions could involve the study of periodic
or random arrays of localized impurities which may be exploited to
induce novel quantum states in Fermi superfluids and the consideration
of a quantum impurity with its own internal degrees of freedom. Such
a system may allow us to explore Kondo physics in cold atoms.

This work is supported by the NSF, the Welch Foundation (Grant No.
C-1669) and by a grant from the Army Research Office with funding
from the DARPA OLE Program. HH acknowledges the support from an ARC
Discovery Project (Grant No. DP0984522). Y.C. was supported by the
NSFC and the State Key Programs of China. We would like to thank
Qiang Han for useful discussions.

\end{document}